# Radio Frequency Performance Projection and Stability Trade-off of h-BN Encapsulated Graphene Field-Effect Transistors

Pedro C. Feijoo, Francisco Pasadas, José M. Iglesias, El Mokhtar Hamham, Raúl Rengel, David Jiménez

P. C. Feijoo, F. Pasadas and D. Jiménez are with the Departament d'Enginyeria Electrònica of the Universitat Autònoma de Bacelona, Cerdanyola del Vallès, 08193 Spain (corresponding author: PedroCarlos.Feijoo@uab.cat).
J. M. Iglesias, E. M. Hamham and R. Rengel are with the Departamento de Física Aplicada, Universidad de Salamanca, Salamanca, 37008 Spain.

**Abstract**

Hexagonal boron nitride (h-BN) encapsulation significantly improves carrier transport in graphene. This work investigates the benefit of implementing the encapsulation technique in graphene field-effect transistors (GFET) in terms of their intrinsic radio frequency (RF) performance, adding the effect of the series resistances at the terminals. For such a purpose, a drift-diffusion self-consistent simulator is prepared to get the GFET electrical characteristics. Both the mobility and saturation velocity are obtained by an ensemble Monte Carlo simulator upon considering the relevant scattering mechanisms that affect carrier transport. RF figures of merit are simulated using an accurate small-signal model. Results reveal that the cutoff frequency could scale up to the physical limit given by the inverse of the transit time. Projected maximum oscillation frequencies, in the order of few THz, are expected to exceed the values demonstrated by InP and Si based RF transistors. The existing trade-off between power gain and stability and the role played by the gate resistance are also studied. High power gain and stability are feasible even if the device is operated far away from current saturation. Finally, the benefits of device unilateralization and the exploitation of the negative differential resistance region to get negative-resistance gain are discussed.

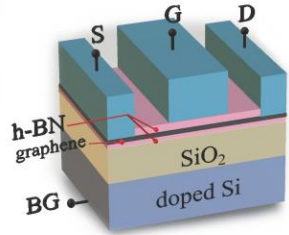

Fig. 1. Schematic of the h-BN encapsulated GFET considered in this work. The terminals are named as S (source), G (gate), BG (back gate) and D (drain).

**INTRODUCTION**

Electron transport properties of graphene improve considerably when this two-dimensional material is encapsulated between two layers of hexagonal boron nitride (h-BN): carrier mobility and saturation velocity can reach experimental values of more than $5 \cdot 10^4$ cm$^2$ V$^{-1}$ s$^{-1}$ and $5 \cdot 10^7$ cm s$^{-1}$, respectively [1]–[3], which gives a significant increase with respect to the values of graphene supported on SiO$_2$ substrates (mobilities of ~4500 cm$^2$ V$^{-1}$ s$^{-1}$ and saturation velocities of ~$10^7$ cm s$^{-1}$ have been reported in [4], [5]). This improvement is caused by the low presence of dangling bonds in the h-BN/graphene interface, the similar chemical structure of graphene and h-BN surface, and the reduced rate of carrier scattering with surface polar phonons [1], [3]. Such a good performance in carrier transport makes h-BN encapsulated graphene an ideal candidate for transistors with applications in power or current amplification of radio frequency (RF) analog signals [6]. The topic of RF performance of graphene field-effect transistors (GFET) has been extensively covered in [7]–[11], including the study of negative differential resistance (NDR) and its effect on RF stability [12], as well as the influence of contact resistance and gate resistance [13], [14].

This work explores the capability of h-BN encapsulated GFETs to deliver power and current gain when used as the active device in microwave/RF amplifiers. For this purpose, the RF figures of merit (FoM) of GFETs are assessed considering the intrinsic properties of the transistor together with the influence of the series resistances at the terminals, ignoring the effects produced by extrinsic capacitances. To compare with our theoretical predictions, experimental data needs a de-embedding procedure to extract the contribution of the parasitic capacitances.

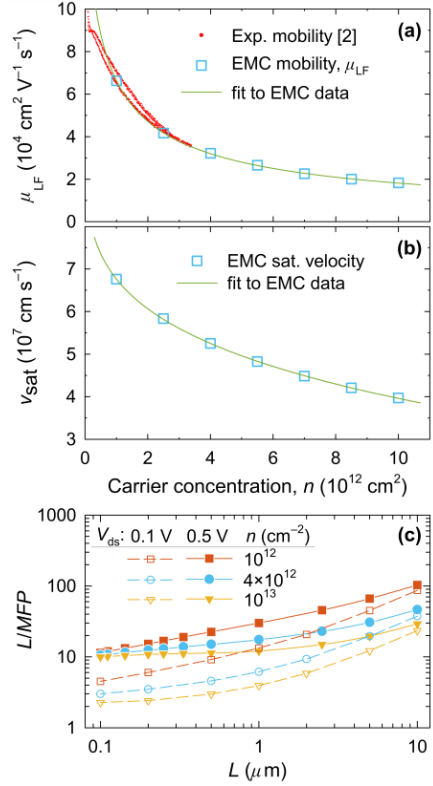

Fig. 2. (a) Electron mobility and (b) velocity saturation calculated by the EMC method as a function of the carrier concentration in the h-BN encapsulated graphene is shown together with the experimental values from [2]. (c) Channel length-MFP ratio as a function of the channel length for $V_{ds}$ of 0.1 and 0.5 V at different carrier concentrations. In all cases, $L$ is larger than MFP, validating drift-diffusion assumption.

**RF PERFORMANCE SIMULATIONS**

We have considered a prototype h-BN encapsulated GFET, the scheme of which is depicted in Fig. 1. The h-BN/graphene/h-BN stack is on top of a thick $SiO_2$ layer, which is, in turn, on top of a highly doped Si wafer acting as the back gate. We have considered 30 nm thick h-BN top and bottom layers with relative permittivity of 3 [15] and a 285 nm $SiO_2$ layer with relative permittivity of 3.9, which are typical values for h-BN encapsulated GFETs [1], [16]. Drain and source series resistances are assumed to have the state-of-the-art values of 100 Ω μm [17]. Narrow width effects can be neglected since channel width has been considered sufficiently large.

Low field carrier mobility $\mu_{LF}$ and carrier saturation velocity $v_{sat}$ in h-BN encapsulated graphene were obtained by a self-consistent ensemble Monte Carlo (EMC) simulator at steady-state conditions that accounts for the scattering mechanisms in this particular encapsulated graphene structure [18]–[20]. The scattering mechanisms include graphene intrinsic phonons, remote surface polar phonons from h-BN encapsulating layers, carrier-carrier collisions and scattering caused by defects and charged impurities. A complete description of the formalism used for each scattering mechanism and the parameters chosen can be found in the supplementary material of [20]. Moreover, the hot phonon effect is also included, which is particularly important at high field conditions in graphene [21]. The levels of impurities and defects have been adjusted so the carrier mobility as a function of the carrier concentration *n* fits the experimental values obtained for h-BN encapsulated graphene in [2]. The values presented here are obtained considering the average velocity for a large number of particles in the simulation and an additional time average once the stationary conditions are reached. The statistical uncertainty is noticeably small: when using five different seeds for the random number generation, the standard error is less than 0.7% in the worst case, which clearly indicates that the numerical error is minimum. A number of particles between $4 \cdot 10^4$ and $4 \cdot 10^5$ have been considered in the simulations, depending on the carrier concentration. This guarantees that the margin of error in the results for a 99% confidence level is less than 0.7%. The results for $\mu_{LF}$ and $v_{sat}$ have been plotted in Fig. 2(a) and (b), where it can be observed that both magnitudes tend to decrease with the carrier concentration, reaching an approximately constant minimum value for carrier concentrations above $10^{13}$ cm$^{-2}$.

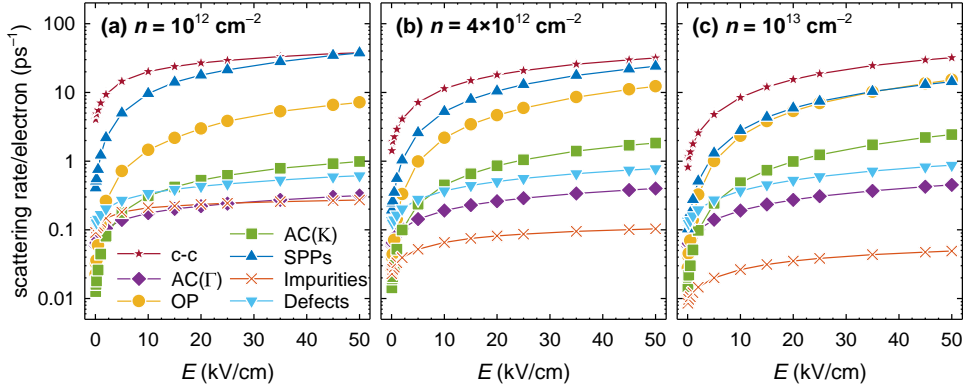

Fig. 3. Average number of scattering events per unit time as a function of the electric field for carrier concentrations of $10^{12}$ cm$^{-2}$ (a), $4 \cdot 10^{12}$ cm$^{-2}$ (b) and $10^{13}$ cm$^{-2}$ (c). The scattering events include carrier-carrier interactions (c-c), acoustic intravalley phonons at the Γ and K points (AC), optical phonons (OP), surface polar phonons from the h-BN layers (SPP), impurities and defects.

Throughout this work we assume that the grain boundaries in graphene are not affecting the carrier transport. This should be valid for monocrystalline graphene and polycrystalline graphene samples with grain size much greater than the channel length. The grain size, in the latter case, is quite sensitive to the growth technique and values of mobility ranging from 2500 to 350,000 cm$^2$ V$^{-1}$ s$^{-1}$ have been reported in the literature; the former case corresponding to all chemical vapor deposited (CVD) h-BN/graphene/h-BN samples [22], and the latter case to transferred CVD graphene encapsulated between exfoliated h-BN layers [2].

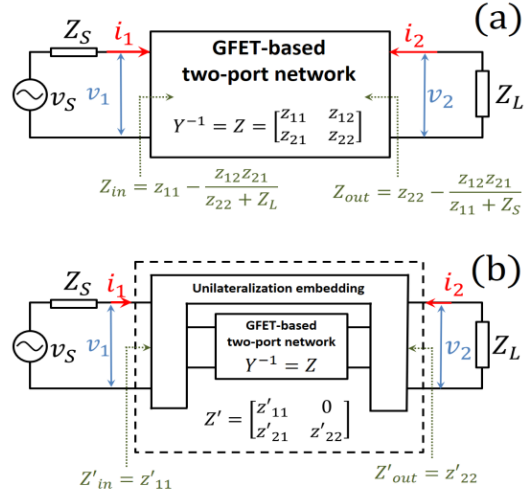

Fig. 4. (a) GFET working as a two-port network, characterized by an impedance matrix $Z$. (b) A four-port passive lossless reciprocal embedding can be added to the GFET for unilateralization, making $z'_{12} = 0$.

Importantly, EMC simulations allow calculation of the mean free path (MFP) and differentiation of the contribution of each scattering mechanism. Fig. 2(c) confirms that the MFP is smaller than the channel length $L$ in all cases of channel length and carrier concentrations in graphene considered in this work, which guarantees the validity of the drift-diffusion mechanisms as the driving forces of the carrier transport. The separate influence of the different scattering mechanism types can be perceived in Fig. 3, where the average number of scattering events suffered by a carrier per unit time is presented as a function of the electric field for several carrier concentrations. Carrier-carrier scattering is the dominant mechanism. However, given its Coulombic nature and the momentum and energy conservation laws for the pair of interacting particles, the influence of each individual collision on the total velocity of colliding carrier pairs is minimal due to the preferred small wavevector transitions, therefore implying a reduced influence on saturation drift velocity and specially in mobility in comparison with other types of mechanisms.

Moreover, at larger carrier concentrations and low fields its rate for each single electron is lower due to the Pauli exclusion principle, that restricts carrier-carrier interactions to those particles close to the Fermi surface [23]. The MFP values obtained are in the same range than in other theoretical studies that take into account electron-electron scattering [24]. At low carrier concentrations, the low-field mobility is strongly influenced by the interactions with the h-BN layers and, to a lesser extent, with defects and impurities, while the saturation velocity is importantly affected by surface polar phonon interactions and intrinsic optical phonons. On the other hand, at high carrier concentrations, defects and surface polar phonons are critical for the low field mobility. The saturation velocity in this high-concentration regime is mostly influenced by intrinsic optical phonons, with a relevant role also of the interactions with the h-BN.

The DC characteristics of the previously described GFET have been simulated by a method that self-consistently solves both Poisson's equation and current-continuity equation [20], [25], introducing in the algorithm $\mu_{LF}$ and $v_{sat}$ data coming from the EMC simulator. This method naturally takes into account the influence of drain voltage $V_{ds}$, which may introduce strong short-channel effects in scaled GFETs. It also includes an appropriate carrier saturation velocity model at high fields [25]. As for the GFET simulation, both $\mu_{LF}$ and $v_{sat}$ have been assumed to be equal for electrons and holes and constant for $n$ below $5 \cdot 10^{11}$ cm$^{-2}$. Throughout this work, we have considered a positive $V_{ds}$, but all the results for channels dominated by electrons or holes, including the cases of ambipolar conduction, can be easily extrapolated to the case of negative $V_{ds}$ by only exchanging the type of carrier.

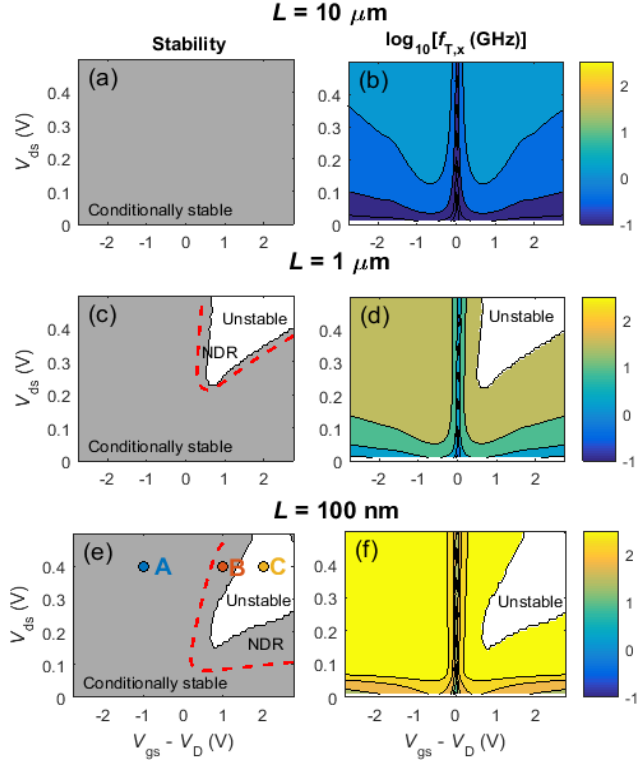

Fig. 5. Stability and extrinsic cutoff frequency ($f_{T,x}$) as a function of the intrinsic bias point for GFETs with $L$ of 10 µm, (a) and (b), 1 µm, (c) and (d) and 100 nm, (e) and (f). Dashed lines in stability graphs correspond to zero intrinsic output conductance $g_d$, which delimit the intrinsic NDR region.

To study the GFET RF performance, we have used the small-signal model determined by the impedance matrix $Z$, the inverse of the admittance matrix $Y$ as described in [20]. For the calculation of the DC charge at each terminal, we have used a linear Ward-Dutton partition, guaranteeing charge conservation in the device [26]. The small-signal model contains information on the transconductance $g_m$ and the output conductance $g_d$ parameters, which are obtained from the resulting perturbation in the drain current that originates from small voltage variations around gate and drain biases, respectively. Additionally, the model includes the $C_{ij}$ transcapacitances, which result from the perturbation observed in the charge of the "i" terminal due to small variations in the voltage of the "j" terminal, where i and j refer to the drain/source/gate terminals. Together with the intrinsic small-signal model of the GFET, the only extrinsic elements that have been taken into account have been the series resistances at source, drain and gate terminals, ignoring the possible effects of parasitic capacitances. This way, current gain, power gain and stability were analyzed considering the GFET as a two-port network, where the top gate - source terminals are acting as the input port, and the drain - source terminals as the output port, with the source being the common terminal [27].

Next, we have simulated the GFET power gain considering a simple amplifier configuration with a voltage source $v_S$ (with internal impedance $Z_S$) connected to the GFET input port and a passive load $Z_L$ connected to the output, as shown in Fig. 4(a). In this configuration we have calculated the maximum available gain (MAG) or the maximum stable gain (MSG) depending on whether the device is stable or potentially stable. The former case implies that the two-port network is stable for all combinations of passive $Z_S$ and $Z_L$, but for the latter only some combinations of passive source an load impedances lead to stability [28]. This configuration is also used to determine the extrinsic cutoff frequency $f_{T,x}$, which corresponds to the frequency at which the extrapolation of low-frequency current gain becomes equal to one.

The examined GFET behaves as a bilateral device at high frequency (beyond 100 GHz) since the impedance parameter $z_{12}$, which is related to the transmission of signals from the output to the input, departs significantly from zero. Because of this, it is interesting to consider a linear lossless reciprocal embedding added to the two-port network as depicted in the schematic of Fig. 4(b) with the purpose of avoiding any reverse signal transmission. The resulting impedance matrix $Z'$ must then verify that its element $z'_{12}$ is null. Maximum power gain calculated in this configuration is the so-called unilateral power gain $U$ [27]. We have determined the maximum oscillation frequency $f_{max}$ as the value at which the extrapolation of $U$ at low frequencies reaches one (0 dB) [28]. The frequencies $f_{T,x}$ and $f_{max}$ are important FoMs in RF applications, especially $f_{max}$, since it is related to the maximum working frequency delivering power gain.

RF stability can be determined from the small-signal model through the $K$-$\Delta$ test [28], which, given a frequency, classifies the device in the configuration shown in Fig. 4(a) into unstable, conditionally stable or unconditionally stable for the selected bias point. A device is stable for a given frequency if $K > 1$ and $|\Delta| < 1$, conditionally stable if $|K|<1$ (or $K > 1$ with $|\Delta| > 1$), and unstable if $K < -1$. As for the unilateralized device, stability is reached when $\text{Re}(z'_{11})$ and $\text{Re}(z'_{22})$ are positive. Otherwise the unilateralized device would behave as conditionally stable.

### RESULTS AND DISCUSSION

Firstly, we are interested in the bias dependent device stability map for the circuit configuration shown in Fig. 4(a). The result has been plotted in Fig. 5(a), (c) and (e), showing the stability behavior versus both the intrinsic $V_{ds}$ and gate voltage $V_{gs}$ for different channel lengths, where we have assumed a representative value of the gate series resistance $R_G$ of 61.6 Ω µm$^2$/$L$ [26], [29]. In the horizontal axis, the gate voltage overdrive has been considered respect to the Dirac voltage $V_D$. At this specific $V_{gs}$, the current $I_{ds}$ exhibits a minimum in the transfer characteristic. It can be estimated as $V_D = V_{gs0} + 0.5\ V_{ds}$, where $V_{gs0}$ is the flatband voltage [29], [30]. We have labelled the device as conditionally stable if $|K| < 1$ for low and medium frequencies and $K > 1$ for high frequencies. On the other hand, the device is labelled as unstable when the stability parameter is $K < -1$ for any frequency range. Notice that even if the device is not operated in the frequency range where $K < -1$, high frequency noise could drive the amplifier to produce undesired oscillations. We have found that, for channel lengths below 1 µm, there exists a region inside the considered bias window where the device is unstable for high frequencies according to the $K$-$\Delta$ test. This instability region appears only at the right part of the $V_{gs}$-$V_{ds}$ map ($V_{gs} > V_D$, the electron branch), inside the NDR region [12], [30], and it grows as the channel becomes shorter. For long-channel devices ($L > 1$ µm) this NDR region would also exist but it is located beyond the computational window. Although the instability region is located inside the NDR area, it is worth noting that, for the short channel transistor (100 nm), there exists a sub-region where the device is conditionally stable under NDR operation. We will comment on the exploitation of this region later on to get negative-resistance gain [31], [32].

To get further insight into the stability for the analyzed amplifier configuration, the $K$-$\Delta$ test of the bias points marked as A, B and C in Fig. 5(e) have been represented in Fig. 6.

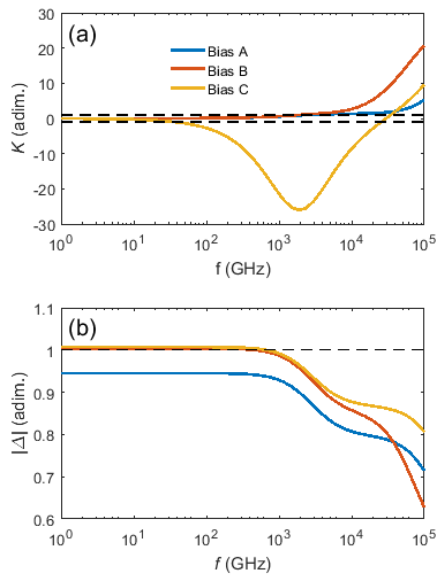

Fig. 6. $K$-$\Delta$ stability test for a h-BN encapsulated 100-nm GFET at different bias points A, B, C corresponding to $V_{gs}$ - $V_D$ = -1, 1 and 2 V, respectively. $V_{ds}$ is 0.4 V in all three cases. Biasing the device at C, values of $K$ are produced below -1 for a range of frequencies, which is indicative of device instability.

Point A is located inside the hole branch ($V_{gs} < V_D$) while B and C belong to the electron branch ($V_{gs} > V_D$), both inside the NDR region. Biasing the device at A or B drives it into conditional stability up to ~ 2 THz and absolute stability beyond that frequency, as shown in Fig. 6(a). However, biasing the device at C produces a range of frequencies where the device is unstable ($K < -1$), so the GFET is unusable in the amplifier configuration analyzed in Fig. 4(a).

Fig. 5(b), (d) and (f) show that $f_{T,x}$, and thus current gain, are symmetric with respect to $V_D$ and increase with channel length reduction. The details on the scalability of the maximum $f_{T,x}$ for different $V_{ds}$ are shown in Fig. 7(a). For long channels, the scaling is proportional to $1/L^2$ but it reaches eventually its physical limit for graphene $v_F/(2\pi L)$, determined by the transit time $v_F/L$, with the Fermi velocity $v_F \sim 10^8$ cm s$^{-1}$. The scaling thus changes to a $1/L$ trend for short channels. It is also noticeable that at channel lengths shorter than 100 nm, $f_{T,x}$ can slightly decrease with $V_{ds}$, which is caused by carrier velocity saturation at the point where the type of carriers changes from electrons to holes inside the channel. According to our simulations, $f_{T,x}$ can reach values up to 5 THz for $L$ down to 30

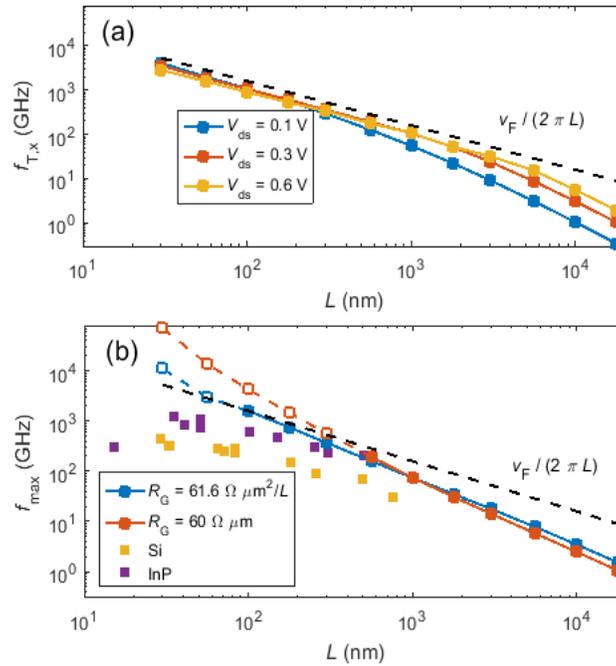

Fig. 7. (a) Scaling of maximum $f_{T,x}$ as a function of channel length $L$ and its dependence with the drain voltage. (b) Scaling of largest possible $f_{max}$ for $V_{ds} = 0.1$ V and $V_{gs} < V_D$ and comparison with state-of-the-art values of RF Si and InP transistors [33]. Blue squares correspond to a $R_G$ upscaling with $L$ reduction and orange squares to a constant $R_G$. Open symbols represent the $f_{max}$ predictions that might be worth of reconsideration under the non-quasi-static hypothesis.

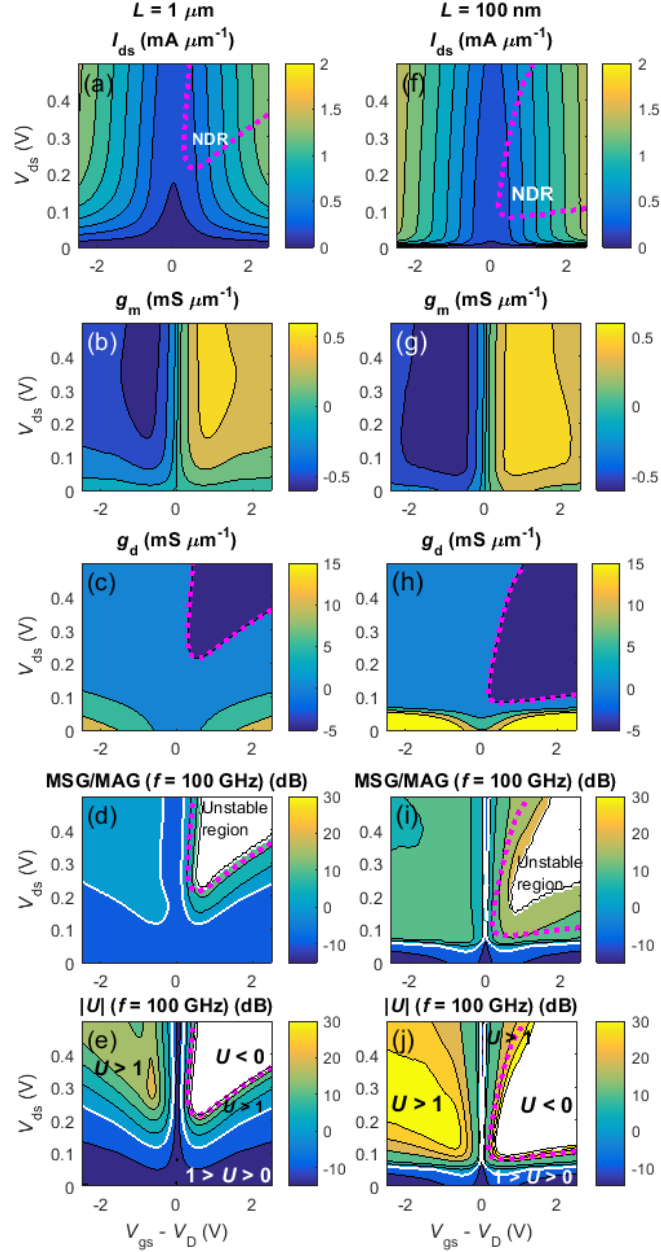

Fig. 8. Drain current $I_{ds}$, intrinsic transconductance $g_m$, output conductance $g_d$, and power gain (MSG/MAG and $U$) at 100 GHz, for a prototype GFET with $L = 1$ μm (a-e) and 100 nm (f-j). Dotted lines indicate the set of points for $g_d = 0$. White regions in the MSG/MAG diagrams correspond to the instability regions. In the $U$ diagrams, white regions correspond to negative values of $U$. White solid lines indicate the set of bias points for which both $U$ and MSG/MAG are equal to one.

nm, a number that is superior to the ones gotten from RF transistors based on Si or InP [33].

Next, we deal with the scaling of $f_{max}$ with the channel length, plotted in Fig. 7(b). The gate bias for $f_{max}$ calculation has been set in the hole dominated $I$-$V$ branch ($V_{gs} < V_D$) in order to maximize its value at $V_{ds} = 0.1$ V. The dashed line in the graph represents the physical limit for the cutoff frequency $v_F/(2\pi L)$ and is represented for comparison. The frequency $f_{max}$ has been calculated for two different scenarios: in the first one, $R_G$ upscales with channel length reduction, and in the second, $R_G$ is kept constant with the channel length [14]. According to our simulations, $f_{max}$ for $L = 30$ nm and $R_G = 2$ kΩ μm can reach values of up to 10 THz. Reducing $R_G$ to 60 Ω μm can improve this value, largely surpassing $v_F/(2\pi L)$. Although $f_{max}$ is not affected by this limit, more accurate calculations may need developing a non-quasi-static small-signal model [34], [35], which is beyond the scope of this work. Nevertheless, our results indicate that GFET transistors based on h-BN encapsulated graphene are expected to overcome the best values of $f_{max}$ demonstrated in technologies based on Si or InP [33].

Both MSG/MAG and $U$ at a working frequency of 100 GHz can be examined in Fig. 8 as a function of the bias, together with $g_m$, $g_d$ and the drain current $I_{ds}$ of the GFET. These magnitudes are represented for channel lengths of 1 μm and 100 nm. In all graphs, the dashed line delimits the NDR region ($g_d < 0$). From Fig. 8 (i), MSG/MAG presents large values when $g_d$ is close to 0 for a channel length of 100 nm: up to 15 dB. However, the proximity of the unstable region could drive the device into an undesired oscillating behavior. Since the NDR sub-region of conditional stability is narrow, it might be preferable to bias the GFET in the hole branch ($V_{gs} < V_D$), keeping the unstable region far from the selected bias point. Although the power gain only reaches 1 dB for the 1 µm channel transistor, the 100 nm long transistor can achieve up to 10 dB in this branch.

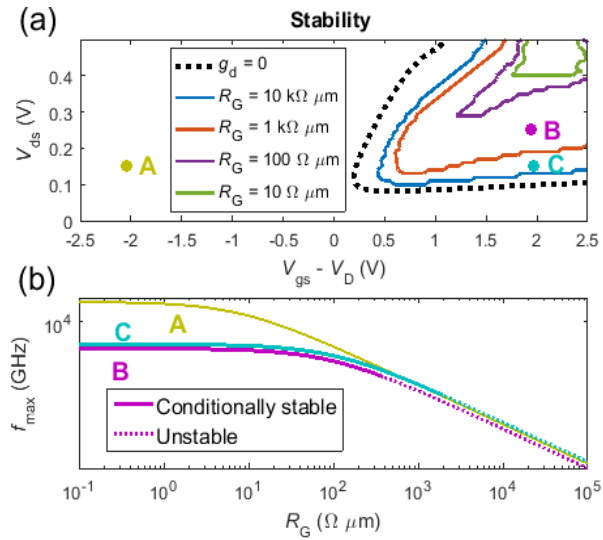

Fig. 9. (a) Instability region for different $R_G$ values. For $R_G < 10$ Ω μm, the area of instability does not further decrease significantly. Dashed line is the limit of the NDR region. (b) $f_{max}$ as a function of the $R_G$ for different bias points marked in (a). When the device is polarized in the electron branch, it becomes unstable for high values of $R_G$.

Regarding the power gain in the unilateralized configuration, it can be observed in Fig. 8(e) and (j) that the GFET presents a wide bias region where unilateral power gain ($U > 1$) can be reached. When the device is biased at the hole branch, the unilateral power gain could be much higher than MSG/MAG for the non-unilateralized case, reaching values of 20 dB and 35 dB for the 1 µm and 100 nm transistors, respectively. This is an unexpected result because the device operates far away from current saturation. In the sub-region of the map where $0 < U < 1$, no power gain amplification is expected. Interestingly, there is a sub-region with $U < 0$ produced when the device is biased in the NDR area. In this sub-region of negative-resistance gain the resulting product Re($z'_{11}$)×Re($z'_{22}$) of the unilateral device is negative [27], [36]. In such a case, the power amplifier needs to be carefully terminated to avoid instability, i.e., it is needed to enforce Re($Z_L + Z'_{out}$) > 0 and Re($Z_S + Z'_{in}$) > 0 in the scheme of Fig. 4(b), where $Z'_{in} = z'_{11}$ and $Z'_{out} = z'_{22}$ [12], [37]. In practice, a careful choice of $Z_L$ and $Z_S$ must be done keeping the unstable region far enough at the expense of a power gain degradation. On the other hand, another interesting option to exploit the $U < 0$ negative-resistance gain sub-region is to use the device in a single-port reflection amplifier configuration, which requires the use of a circulator. Since the real part of $z'_{22}$ (or $z'_{11}$) is negative, the device presents power gain in port 2 (or port 1). The NDR device can send back an amplified signal to the circulator, which would separate incident and reflected waves [31], [38].

Let us focus now on the role that $R_G$ plays in defining the stability and power gain. To illustrate this, we have plotted in Fig. 9(a) the stability map corresponding to a 100 nm GFET with $R_G$ ranging from 10 kΩ µm down to 10 Ω µm. The dotted line defines the edge of the NDR region, while the solid lines define the stability region edge. As the value of $R_G$ decreases, the area of instability in the map is reduced. Further decrease of $R_G$ below 10 Ω µm does not reduce significantly the region of instability. The choice of the bias point is hence important to guarantee a proper amplifier design. Regarding the maximum operating frequency with power gain, Fig. 9(b) illustrates the behavior of $f_{max}$ as a function of $R_G$ for different biases, which are marked in Fig. 9(a). In the graph, the dotted curves indicate that the device becomes unstable in these conditions. As $R_G$ grows, $f_{max}$ falls to zero in all cases. The GFET remains conditionally stable if it is biased in the hole branch (point A, $g_d > 0$) while it turns from conditionally stable into unstable if biased in the electron branch (point B, $g_d < 0$). In the latter case, as the $R_G$ increases, input and output ports get progressively decoupled and the value of the output impedance $Z_{out}$ approaches $Z_{22}$. The real part of $Z_{22}$ is negative due to the NDR, so $\text{Re}(Z_{out}) < 0$ may cause oscillations at the output. If a lower $V_{ds}$ is chosen, the limit of stability appears for a larger $R_G$. All of this leads to the conclusion that it is quite significant to have a low $R_G$ in order to both minimize the unstable region and maximize $f_{max}$ for a particular bias point.

## CONCLUSIONS

In this work, we have first simulated the carrier mobility and saturation velocity in h-BN encapsulated graphene via an accurate EMC simulator that takes into account the complex interplay among the scattering mechanisms impacting on the carrier transport and correctly reproduces experimental mobility measurements. The results are fed into a self-consistent drift-diffusion simulator, whose output are the DC characteristics of the h-BN encapsulated GFET. Then, the transistor RF FoMs together with the stability map have been obtained by using an appropriate intrinsic small-signal model of the GFET, which includes the non-reciprocal capacitances of the GFET, along with the series resistances at source, drain and gate. Parasitic capacitances have not been considered. We have found that channel length scaling is beneficial in terms of RF performance although there exists a trade-off between power gain and device stability. Specifically, we have found that $f_{T,x}$ could be scaled up to the physical limit imposed by the inverse of the transit time, and can reach several THz. It can also be expected that $f_{max}$ will reach the THz range for channel lengths below 100 nm, being h-BN encapsulated GFETs potentially superior to InP and Si RF transistors. We have found that the device instability is related to the NDR operation. However, biasing the device at the hole branch, even far from current saturation, both high power gain and stability are feasible. We have also discussed the benefits of device unilateralization to get power gain and negative-resistance gain. Finally, we have quantified the relevance of $R_G$ reduction in getting both maximum RF performance and device stability.

**ACKNOWLEDGMENT**

Authors would like to thank Prof. Francisco García Ruiz for the helpful discussions.

This work is funded by the European Union's Horizon 2020 research and innovation program under grants agreements No GrapheneCore1 696656 and No GrapheneCore2 785219, the Generalitat de Catalunya (2014 SGR 384), the Junta de Castilla y León (PhD grant SA176-15) and the Ministerio de Economía y Competitividad (TEC2016-80839-P, TEC2015-67462-C2-1-R, MINECO/FEDER and FJCI-2014-19643).